\documentclass{optica-article}

\journal{opticajournal} 

\articletype{Research Article}

\usepackage{lineno}
\usepackage{siunitx}
\usepackage{booktabs}
\usepackage{multirow}

\begin{document}

\title{Mechanical loss and stability analysis of NEXCERA in ultra-stable optical cavities}

\author{Nico Wagner,\authormark{1,2,$\dag$,*} Mateusz Naro\.znik,\authormark{3,$\dag$} Marcin Bober,\authormark{3} Steffen Sauer,\authormark{1,2} Micha\l{} Zawada,\authormark{3} and Stefanie Kroker\authormark{1,2,4}}

\address{
\authormark{1}Institut für Halbleitertechnik, Technische Universität Braunschweig, Hans-Sommer-Str. 66, 38106 Braunschweig, Germany\\
\authormark{2}Laboratory for Emerging Nanometrology, Langer Kamp 6a-b, 38106 Braunschweig, Germany\\
\authormark{3}Institute of Physics, Faculty of Physics, Astronomy and Informatics, Nicolaus Copernicus University, Grudzi\c{a}dzka 5, PL-87-100 Toru\'n, Poland\\
\authormark{4}Physikalisch-Technische Bundesanstalt, Bundesallee 100, 38116 Braunschweig, Germany\\
\authormark{$\dag$}The authors contributed equally to this work.
}

\email{\authormark{*}nico.wagner@tu-braunschweig.de}

\begin{abstract*} 
NEXCERA has emerged as a ceramic-based material for spacers in ultra-stable optical cavities, with a coefficient of thermal expansion that crosses zero near room temperature.
In such cavities, frequency stability is ultimately limited by Brownian thermal noise in the cavity components.
A key parameter in this context is the mechanical loss, which has remained unknown for NEXCERA.
In this work, we investigate the mechanical loss of NEXCERA N117B at room temperature for various resonances using the gentle nodal suspension technique.
We measure a promising minimum mechanical loss of $\phi = \num{1.89e-5}$, indicating the suitability of NEXCERA for low-noise optical cavities.
Using this value, we calculate the thermal noise of a cavity with a NEXCERA spacer and compare its performance to established materials such as ULE and Zerodur, taking into account different mirror substrate options.
Our analysis shows that NEXCERA is a strong candidate for ultra-stable cavities due to its low thermal noise. Combined with its previously reported low linear drift, it offers a highly attractive option for long-term stable optical frequency references.
\end{abstract*}

\section{\label{sec:introduction}Introduction}

Ultra-stable optical cavities rank among the most precise measurement instruments available today, with applications in optical lattice clocks~\cite{Ludlow_2015, Matei_2017, Zhang_2017, Oelker2019}, gravitational wave detection~\cite{Aso_2013, abbott2016observation, Sathyaprakash2012, Abac2025}, and fundamental physics experiments~\cite{Ashby2003, Roberts2017, Delva2018, Milner2019, Kennedy2020}, including tests of general relativity and probes of fundamental constants.
The sensitivity of these systems is ultimately limited by various noise sources. In optical cavities, this includes seismic vibrations, vacuum fluctuations, temperature instabilities, residual amplitude modulation, and parasitic etalons. Many of these can be mitigated through careful system design, incorporating isolation techniques, vacuum engineering, and optimized optical and electronic components~\cite{Nazarova_2006, Chen_2006, Braccinni_2005, Riehle_2003, Legero_2010}.

Among all fundamental limitations, Brownian thermal noise in the cavity components, i.e. mirror substrates, coatings, and the spacer, currently defines the noise floor of state-of-the-art cavities~\cite{Matei_2017, numata2004thermal, kessler2012sub, hafner20158}.
Mitigation strategies have included using crystalline substrates~\cite{Nawrodt2007, Nawrodt2008b}, optimizing coating materials~\cite{Penn_2003, Martin_2010, Martin_2008, Harry_2007, Granata_2020}, and exploring novel coating technologies~\cite{Crooks_2006, cole2013tenfold, Vajente_2021, Dickmann_2018, McGhee_2023, dickmann2023experimental}.
Further improvements have been achieved by operating at cryogenic temperatures~\cite{Matei_2017, Zhang_2017}, enlarging the beam spot size~\cite{Amairi_2013}, or using higher-order and non-Gaussian modes~\cite{Vinet_2010, Bondarescu_2006, Ambrosio_2003}.

In addition to the mirrors and coatings, the spacer contributes  to the thermal noise budget. Beyond maintaining mirror separation, it plays a crucial role in ensuring long-term frequency stability by minimizing drift.
To meet these requirements, two materials are widely used: Ultra-Low Expansion (ULE) glass from Corning Inc. and Zerodur\textsuperscript{\textregistered} from Schott AG, both valued for their extremely low coefficients of thermal expansion (CTEs) near room temperature. 

NEXCERA\textsuperscript{\texttrademark}, a ceramic developed by Krosaki Harima, has gained attention for its ultra-low thermal expansion and high stiffness, promising reduced thermal noise and improved long-term stability. With a Young’s modulus of \SI{140}{\GPa}, substantially higher than ULE (\SI{67.6}{\GPa}) and Zerodur (\SI{90.3}{\GPa}), NEXCERA is stiffer and therefore more resistant to vibrations, making it a strong candidate for next-generation spacer materials in ultra-stable optical cavities~\cite{NipponSteel2001, Barbarat2025}.

Assessing NEXCERA’s performance requires measuring its mechanical loss, which directly determines the thermal noise and thus the achievable frequency stability. Previous work reported a minimum Allan deviation of $\sigma_{y} \approx \num{e-15}$ for NEXCERA-based cavities~\cite{Ito2017}, still an order of magnitude above the coating-limited thermal noise floor of state-of-the-art ULE cavities~\cite{Matei_2017, Robinson_2021}.

Although the thermal noise floor of state-of-the-art cavities is typically dominated by mirror substrates and coatings, demonstrating that the thermal noise contribution of the spacer itself is lower than that of the mirrors is an important step.
In addition, NEXCERA’s lower drift rates and higher stiffness further enhance its suitability for ultra-stable cavities. These combined properties of low thermal noise, reduced drift, and mechanical robustness make NEXCERA a highly promising candidate for next-generation precision experiments requiring long integration times or absolute frequency references, including dark matter searches~\cite{Wcislo_2018}, tests of Lorentz invariance~\cite{Chen_2016, Gurlebeck_2018}, constraints on quantum fluctuations~\cite{Chen_2016, Schiller_2004}, and searches for particles beyond the Standard Model~\cite{Naroznik_2023, Geraci_2019}.

The article is organized as follows. Section~\ref{sec:mechanical_loss_NEXCERA} presents the first measurement of the mechanical loss of NEXCERA N117B at room temperature.
Section~\ref{sec:noise} estimates the thermal noise performance of a cavity with a NEXCERA N117B spacer. Using the measured loss factors and the fluctuation–dissipation theorem, we calculate the Brownian noise and compare it to cavities with conventional spacer materials. Finally, we propose a cavity design based on a NEXCERA N117B spacer and show that fused silica mirror substrates yield better thermal noise performance than ULE substrates.

\section{\label{sec:mechanical_loss_NEXCERA}Mechanical loss of NEXCERA}

Mechanical loss measurements typically involve exciting a material to mechanical resonance and assessing its ringdown time or spectral linewidth~\cite{Reid2016}. For ultra-low-loss materials, these measurements often suffer from significant uncertainties due to excess losses introduced by the suspension system~\cite{Nawrodt2007}. Such issues are particularly pronounced at room temperature, where suspension-related and environmental dissipation mechanisms are more prominent~\cite{Cesarini2009}. Accurate measurement of intrinsic mechanical losses is crucial for evaluating Brownian thermal noise and thus requires careful mitigation of these extrinsic effects. For this purpose, Cesarini \textit{et al.}~\cite{Cesarini2009} developed the gentle nodal suspension technique.
The core principle of this technique is to suspend the sample at a single point, maintaining stable equilibrium through gravity.

In this section, we investigate three NEXCERA disks of type N117B. All disks are geometrically identical, with a diameter of \SI{50.077+-0.007}{\milli\meter}, a thickness of \SI{0.5427+-0.0021}{\milli\meter}, and a surface roughness of $\SI{1.53+-0.16}{\nano\meter}$.

\subsection{\label{sec:experimental}Experimental setup}

\begin{figure}[t]
	\centering
	\includegraphics[width=0.7\linewidth]{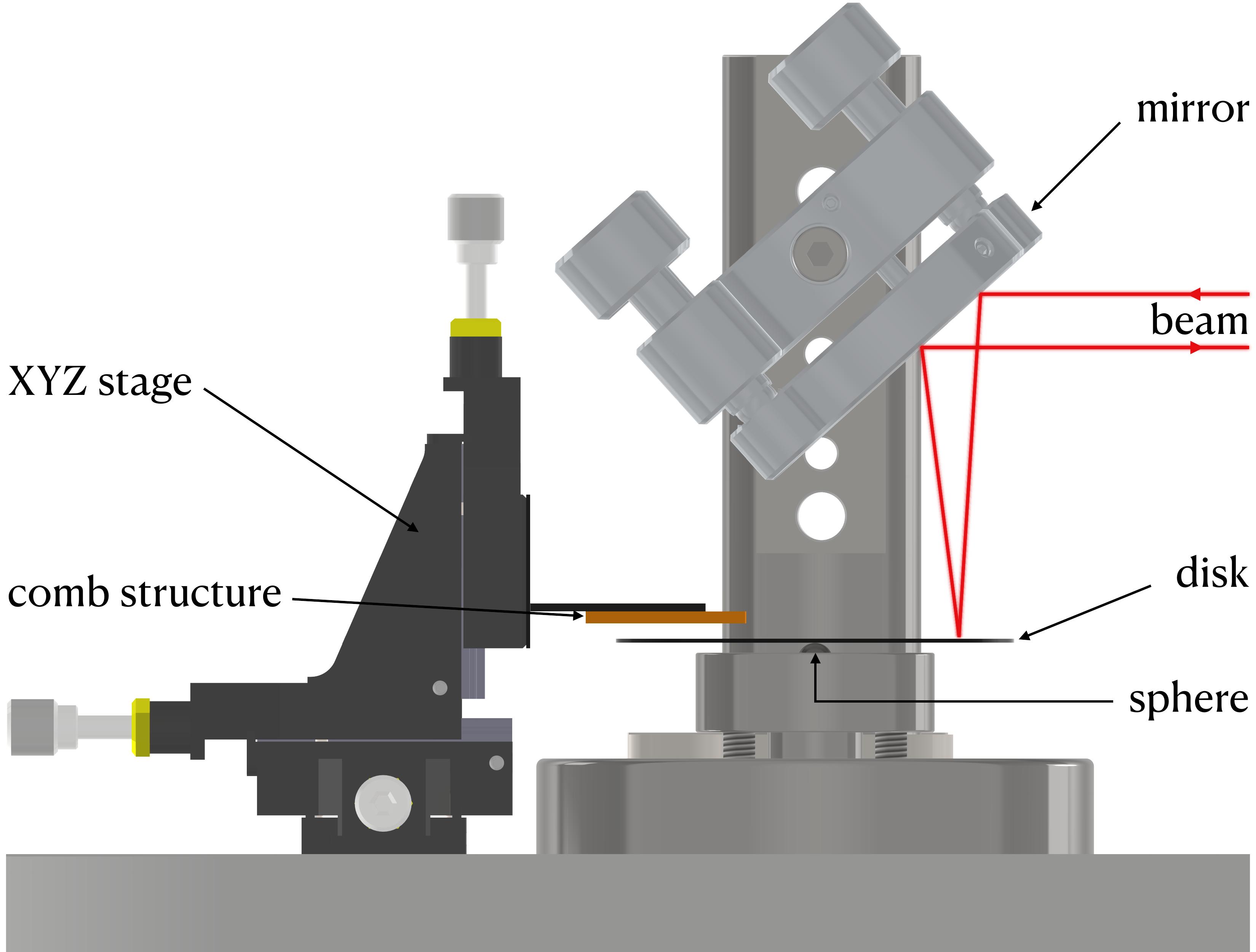}
	\caption{Side view of the nodal suspension setup for mechanical loss measurements. The disk is suspended on a \SI{4.5}{\milli\meter} steel sphere due to gravity. A comb-shaped electrode, used to excite the disk via a high-voltage amplifier, is positioned close to the disk using an XYZ stage. For optical readout, a mirror is mounted at an angle of \SI{45}{\degree} to reflect the incident and reflected laser beams out of the vacuum chamber.}
	\label{fig:sideview}
\end{figure}

Figure~\ref{fig:sideview} shows as a detailed view of the nodal suspension system.
To minimize residual gas damping and its impact on the measured mechanical loss, we operate the experiment under high-vacuum conditions. A two-step pumping process is used: a rotary vane pump performs the initial evacuation of the chamber, followed by a turbomolecular pump mounted directly on top of the vacuum cap to maintain pressures down to \SI{e-7}{\milli\bar}. Vibrations from the turbomolecular pump, operating at tens of thousands of revolutions per minute, are at frequencies high enough to be effectively damped by the mass of vacuum chamber.

We use a custom disk suspension based on the design by Cesarini \textit{et al.}~\cite{Cesarini2009, Cesarini2014}, in which a \SI{4.5}{\milli\meter} steel sphere is rigidly clamped by a conical head structure and protrudes approximately \SI{1}{\milli\meter} above the surface. The stability of the disk suspension depends on both the geometry of the disk and the mechanical properties of the spherical support~\cite{Cesarini2009,Vajente2017}.

For the geometries used in our setup, this yields a stability range of $\lessapprox \pm \SI{17}{\degree}$, which is well above the maximum oscillation amplitude of the disk. Although the influence of the suspension point on the mechanical loss is relatively small~\cite{Cesarini2009}, minimizing this contribution is essential for approaching the intrinsic mechanical loss. To reduce this effect, we adjusted the suspension point using an optical lever technique in which a \SI{45}{\degree} mirror directs the laser beam onto the disk surface, and the reflected beam is used to monitor relative displacement. By shifting the suspension point to a new equilibrium and comparing the reflected spot with its original position, we estimate the disk’s displacement with micrometer accuracy.
The disks are manually placed and positioned on the sphere. Once a disk is suspended in stable equilibrium, its position can be changed by gently poking the edges of the disk with tweezers. Additionally, the position of the suspended disk inside the vacuum chamber can also be altered by tapping the chamber wall with a rubber hammer.

\begin{figure}[t]
	\centering
	\includegraphics[width=0.75\linewidth]{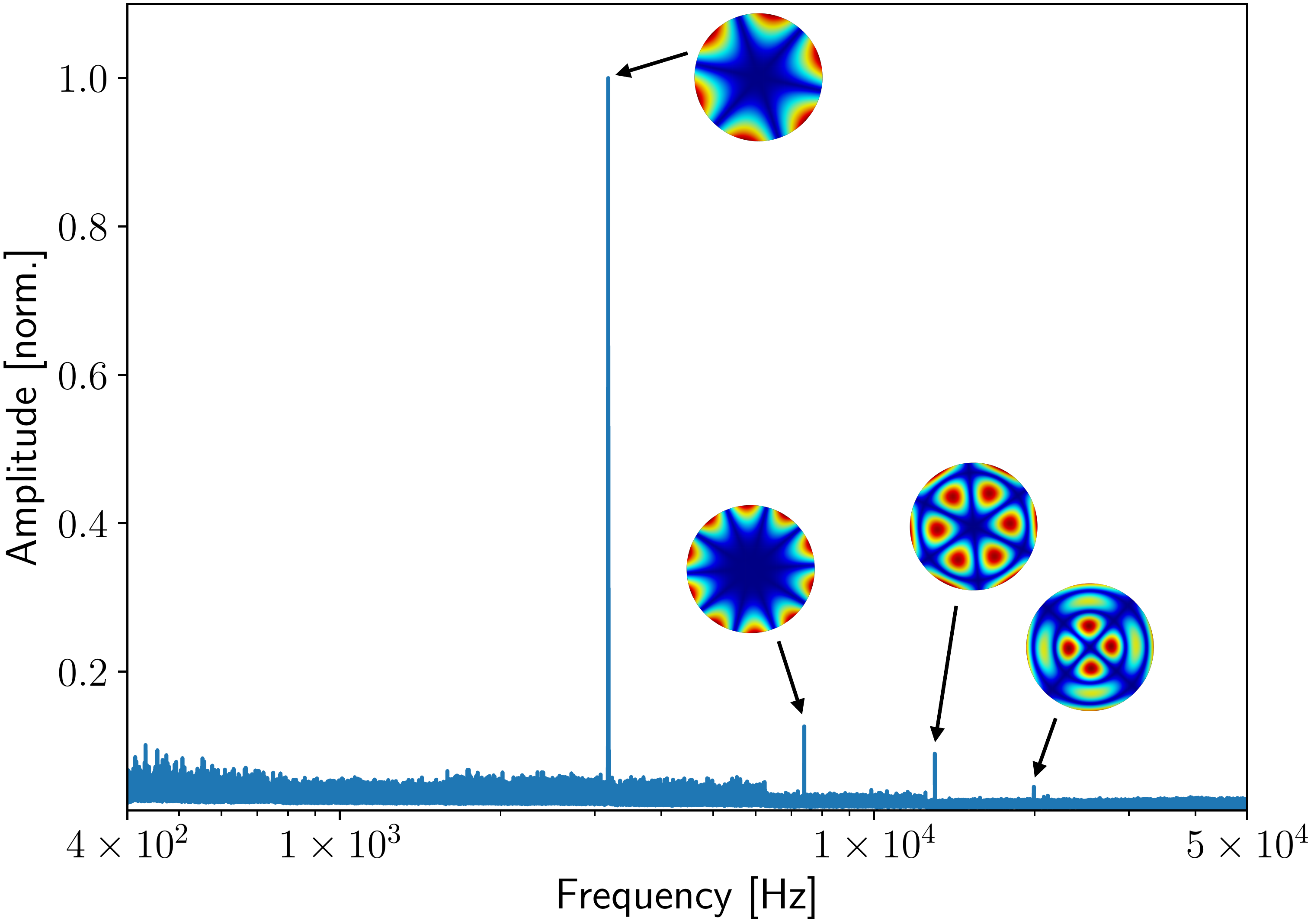}
	\caption{Frequency scan of the mechanical resonances from \SI{400}{\Hz} to \SI{50}{\kHz} with the corresponding mode pattern simulated with COMSOL. Four different mechanical modes can be observed, which are used for further mechanical loss investigations.
	The measured rms voltages of the frequency scan are normalized to the maximum rms voltage recorded by the quadrant photodetector in this measurement.}
	\label{fig:spectrum}
\end{figure}

Furthermore, we estimate the surface area according to Ref.~\cite{Cesarini2009}.
In our configuration, using a \SI{4.5}{\milli\meter} sphere, the contact radius is approximately \SI{8}{\micro\meter}. Therefore, additional losses due to the suspension are expected to be negligible~\cite{Vajente2017}. To minimize the potential influence of surface roughness and defects on the measured mechanical loss, we use steel spheres of quality class G3 according to DIN 5401 standards~\cite{DIN5401}. This class ensures high dimensional and shape accuracy, as well as excellent surface finish.

The disks are excited using a non-contact comb structure (see Fig.~\ref{fig:sideview}) positioned approximately \SI{2}{\milli\meter} above the surface. 
An alternating electric field applied across the capacitor drives the mechanical eigenmodes. 
For this purpose, a Keysight 33500B waveform generator supplies the signal to a high-voltage amplifier (up to $V_{\mathrm{p}} = \SI{800}{\V}$). 
The generator provides a frequency resolution of \SI{1}{\micro\Hz} and an absolute accuracy of $\pm 0.1~\mathrm{ppm}$, ensuring that its frequency stability does not influence the determination of the mechanical resonance linewidths. 

The resulting oscillations are detected optically using a \SI{635}{\nano\meter} laser diode. 
The laser beam is directed onto the disk surface via a mirror, and the reflected light exits the vacuum chamber at a small angle. 
A steering mirror and lens system focus the reflected beam onto the center of a quadrant photodetector (QPD), which senses angular deviations induced by the disk motion. 
From the QPD signal, both oscillation frequency and amplitude can be extracted. 
The detector output is digitized using a National Instruments\textsuperscript{\texttrademark} NI USB-6363 data acquisition card with a maximum sampling rate of \SI{400}{\kHz}. 

Resonance spectra are obtained using a Python-controlled frequency sweep. 
The drive frequency range is divided into small increments, and at each step the disk is excited and the rms signal from the QPD is recorded, providing the oscillation amplitude in volts. 
To avoid peak distortion from residual oscillations, an appropriate settling time is introduced between frequency steps.

\begin{figure}[t]
	\centering
	\includegraphics[width=0.75\linewidth]{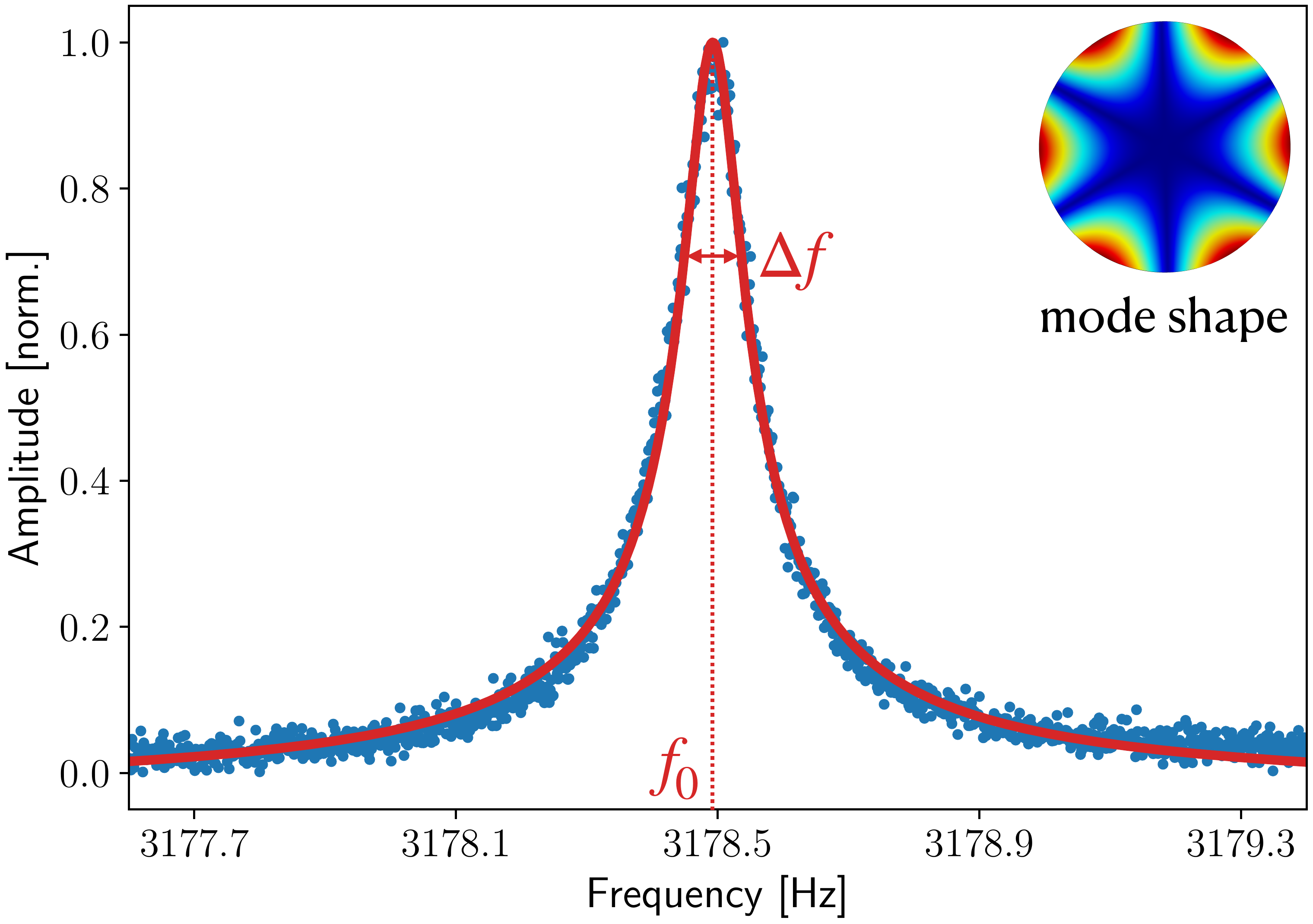}
	\caption{Exemplary frequency scan of a resonance near \SI{3178.5}{\Hz}. The vertical axis shows the normalized oscillation amplitude measured as rms voltages at the quadrant photodetector. The resonance frequency $f_0$ and linewidth $\Delta f$ are used to determine the mechanical loss $\phi$. The inset qualitatively depicts the mode shape, with blue areas indicating nodes (low oscillation amplitude) and red areas indicating antinodes (high oscillation amplitude).}
	\label{fig:resonance}
\end{figure}

\subsection{\label{sec:results}Measurement results}

Figure~\ref{fig:spectrum} shows a frequency scan from \SI{400}{\Hz} to \SI{50}{\kHz} to identify the disk's resonances. We observe four resonances at frequencies of $f_1 = \SI{3.179}{\kHz}$, $f_2 = \SI{7.397}{\kHz}$, $f_3 = \SI{12.99}{\kHz}$, and $f_4 = \SI{21.16}{\kHz}$, respectively. The corresponding mode shapes, obtained from finite element simulations in COMSOL Multiphysics~\cite{Comsol}, are also displayed in Fig.~\ref{fig:spectrum}.

While the simulated mode sequence agrees with the experiment, the absolute values of the measured eigenfrequencies deviate from the simulation results. 
Since the sample thickness strongly affects the resonance frequencies, we adjusted this parameter so that the first simulated mode matches the first measured resonance, following Ref.~\cite{Vajente_2021}. 
The higher-order simulated resonances are then $f_2 = \SI{8.54}{\kHz}$, $f_3 = \SI{13.51}{\kHz}$, and $f_4 = \SI{21.51}{\kHz}$. 
Discrepancies of a few hundred hertz up to about \SI{1}{\kHz} between the simulated and measured eigenfrequencies were also observed in other mechanical loss investigations~\cite{NawrodtPhD}, as the simulations assume idealized free disks with nominal geometry and literature material parameters.
In practice, small uncertainties in thickness, diameter, or suspension conditions, as well as variations in material properties (e.g. Young’s modulus, density, surface layers, or possible anisotropy and inhomogeneity), can shift the resonance frequencies. 
We therefore use the COMSOL results primarily to identify mode shapes and relative spacing rather than as exact predictions of the absolute frequencies.

The mechanical loss is determined from the spectral linewidth of each resonance~\cite{Hofer2010}, obtained by fitting the experimental data with a Lorentzian function, as illustrated in Fig.~\ref{fig:resonance}. It is given by
\begin{equation}
    \phi = \frac{\Delta f}{f_0} \, ,
\end{equation}
where $\Delta f$ is bandwidth in which the signal amplitude drops to a factor of $1/\sqrt{2}$ and $f_0$ is the resonance frequency.
To verify that the measurement signal is independent of the excitation voltage, we measure the resonance peaks over a range from \SI{200}{\V} to \SI{800}{\V} (see Fig.~S1 in Supplement 1). The measured linewidth remains constant within this range.

\begin{figure}[tb]
	\centering
	\includegraphics[width=0.75\linewidth]{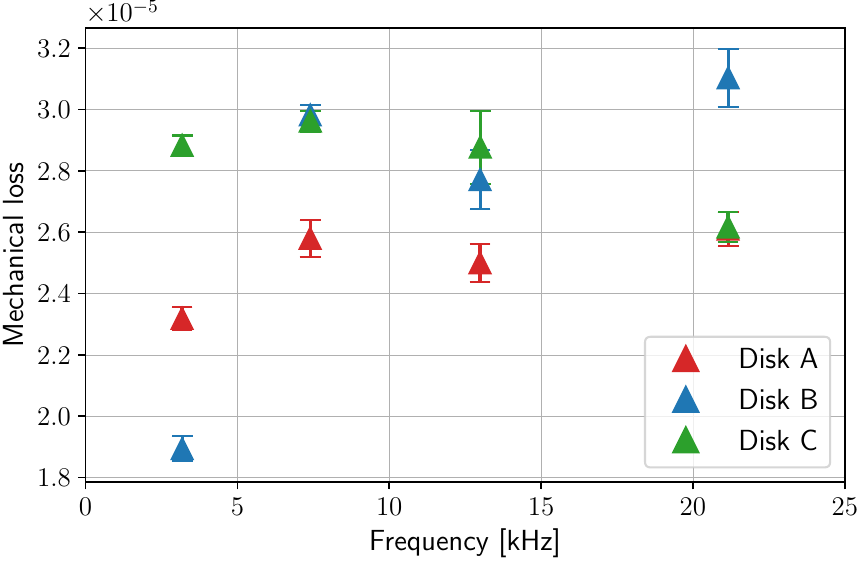}
	\caption{Mechanical loss results for all three investigated disks and four different resonances. The lowest measured mechanical loss is $\phi=\num{1.894+-0.041e-5}$ at $f_0=\SI{3.2}{\kHz}$.}
	\label{fig:losses_disks}
\end{figure}

Figure~\ref{fig:losses_disks} shows the measured mechanical losses for all disks and observed resonances. The lowest loss is found to be $\phi = \num{1.894(41)e-5}$. Error bars represent the square roots of the diagonal elements of the covariance matrix from the fit, reflecting uncertainties in the fitted parameters.

There is up to a \SI{36}{\percent} variation in mechanical loss across different resonances and disks. Such variations are typical, as mechanical loss measurements are known to exhibit significant scatter~\cite{Vajente2017, Crooks2004, Silenzi2024}. Among possible contributing factors, the surface roughness shows a variation of about \SI{10}{\percent} between disks. Even assuming that surface loss accounts for up to \SI{30}{\percent} of the total loss~\cite{Numata2000, Nawrodt2013}, this would translate to less than a \SI{5}{\percent} change in the overall loss. Thus, surface roughness is unlikely to account for the observed differences in mechanical loss.

Due to NEXCERA’s low CTE, thermoelastic losses, which typically dominate in crystalline materials such as silicon at room temperature~\cite{Nawrodt2013}, are not a limiting factor. Using the computation outlined in Refs.~\cite{Lifshitz1999,Fejer2004,Norris2005,Cagnoli2018,Heinert2010} we utilize COMSOL Multiphysics to numerically calculate the thermoelastic losses to be in the range of \num{1e-10} and less. These values are substantially smaller than our measured results. 

Furthermore, we checked the influence of the suspension by performing repeated measurements with slight tilts of the disk on the sphere, as described in Sec.~\ref{sec:experimental}. This resulted in a relative increase in the loss angle of up to \SI{10}{\percent}, consistent with the observations in Ref.~\cite{Cesarini2009}.

\begin{figure}[tb]
	\centering
	\includegraphics[width=0.75\linewidth]{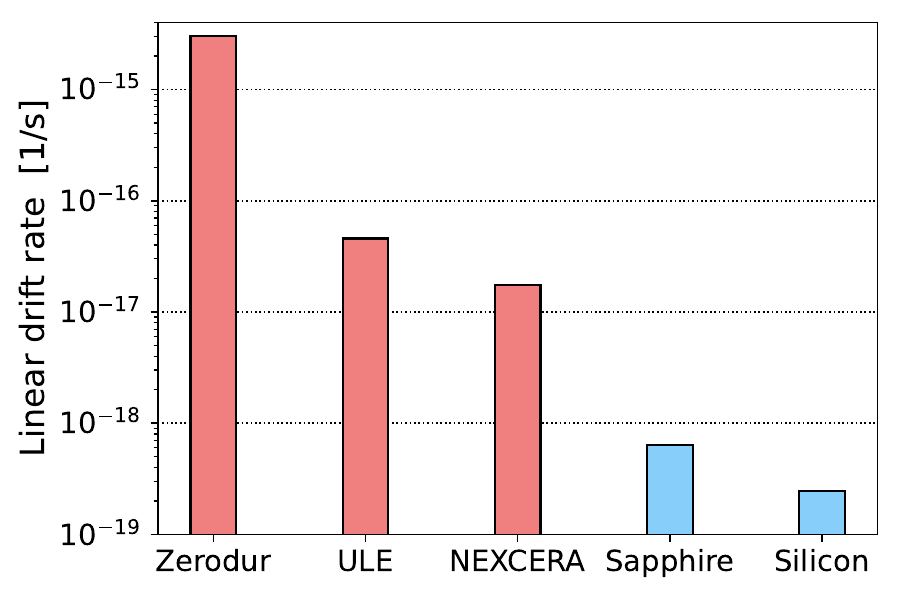}
	\caption{Drift rate for the most common materials used for the ultra-stable optical resonator spacers. Zerodur~\cite{Keupp2005}, ULE~\cite{Keupp2005} and NEXCERA N117B~\cite{Ito2017} operate at room temperature, whereas the single-crystal silicon~\cite{Oelker2019} and sapphire~\cite{Wiens2016} are used in a cryogenic environment. It is important to note that ULE drift rates vary significantly, with reported values ranging from \SI{2.041e-16}{\per\second}~\cite{Alnis_2008} to \SI{4.572e-17}{\per\second}~\cite{Keupp2005}. This variation may be due to differences in measurement duration, environmental conditions, and specific glass samples.}
	\label{fig:drift_rate}
\end{figure}

\section{Performance projection of the NEXCERA cavity}\label{sec:noise}

In the following sections, we present the projected performance of NEXCERA as a cavity spacer material and compare it to other common materials, such as Zerodur and ULE. We evaluate drift rate, Brownian thermal noise, and thermoelastic noise. Additionally, we discuss the noise improvement in NEXCERA cavities using fused silica mirror substrates compared to previously used ULE substrates.

\subsection{\label{sec:Drift rate}Drift rate}

Common spacer materials used at room temperature experience significant frequency drifts due to the crystallization of their amorphous structure. This aging process causes the length of a Fabry-Perot spacer made of ULE, Zerodur, NEXCERA, or similar glass-ceramics to decrease, resulting in a positive drift in resonance frequency.
Figure~\ref{fig:drift_rate} compares the frequency drift rates of commonly used spacer materials, divided into amorphous materials at room temperature (red bars) and crystalline materials at cryogenic temperatures (blue bars).
Both, ULE and NEXCERA exhibit drift rates two orders of magnitude lower than Zerodur, with NEXCERA N117B having a drift rate of \SI{1.738e-17}{\per\second} compared to the best-reported value for ULE at \SI{4.572e-17}{\per\second}~\cite{Keupp2005}.
Aging processes are known to slow down over time as the material structure gradually stabilizes~\cite{Phillips_1996, Dube_2009}. This progressive stabilization leads to a decreasing drift rate, suggesting that older materials may exhibit improved long-term performance. According to Fig.~\ref{fig:drift_rate}, the accumulated drift of NEXCERA and ULE spacers reaches the cavity thermal noise limit within a few seconds.

As shown in Fig.~\ref{fig:drift_rate}, the drift rate can be greatly reduced using crystalline materials such as single-crystal silicon or sapphire, which exhibit values nearly two orders of magnitude lower than NEXCERA.
However, both silicon~\cite{Middelmann_2015, Wiens_2020} and sapphire~\cite{Touloukian_1977, Taylor_1996} have CTEs two to three orders of magnitude higher than ULE or NEXCERA at room temperature, making them significantly more sensitive to temperature changes. This restricts the operational temperature of silicon to the zero-crossing at \SI{16}{\K} and \SI{124}{\K}~\cite{Lyon1977}. 
Furthermore, since the CTE approaches zero near \SI{0}{\K}~\cite{Roset2025}, sapphire and silicon cavities often operate at ultra-low temperatures around \SI{4}{\K}~\cite{Zhang_2017, Muller_2003}, \SI{1.5}{\K}~\cite{Wiens_2014, Schiller_1997, Wiens_2023}, or even in the sub-Kelvin regime~\cite{Barbarat_2023}. Despite their crystalline structures being less susceptible to aging effects, silicon and sapphire cavities still exhibit small long-term drifts in practice~\cite{Wiens_2020, Hartnett_2006}.

Currently, no material combines both an ultra-low CTE at room temperature and a crystalline structure. NEXCERA is promising candidate for low-drift cavity spacers in room-temperature applications, provided its thermal noise proves sufficiently low.

\subsection{Noise performance and stability limits of NEXCERA spacers}

In this section, we present the computed Brownian thermal noise and thermoelastic noise.
The noise calculations are based on the fluctuation-dissipation theorem~\cite{Callen_1951, Callen_1952}.
Details on the computation and utilized material parameters can be found in Supplement~1.

The mechanical loss presented in Sec.~\ref{sec:results} enables the estimation of the Brownian thermal noise for a spacer made of NEXCERA N117B. For this projection, we use the lowest measured mechanical loss, $\phi = \num{1.89e-5}$, as it is least affected by parasitic losses and thus best represents the intrinsic loss relevant to Brownian thermal noise. Samples with high surface-to-volume ratios typically exhibit larger mechanical losses due to additional surface-related mechanisms~\cite{Nawrodt2013, Nawrodt_2008, Penn2006}. In contrast, cavity spacers have much lower surface-to-volume ratios than the disks studied here. Therefore, using our lowest measured loss provides a conservative estimate for Brownian thermal noise. We furthermore assume a frequency-independent loss angle, as no significant frequency dependence was observed in the measured mechanical losses across the kHz range. Accordingly, the structural damping model is adopted in our thermal noise analysis.

\begin{figure}[tb]
	\centering
	\includegraphics[width=\linewidth]{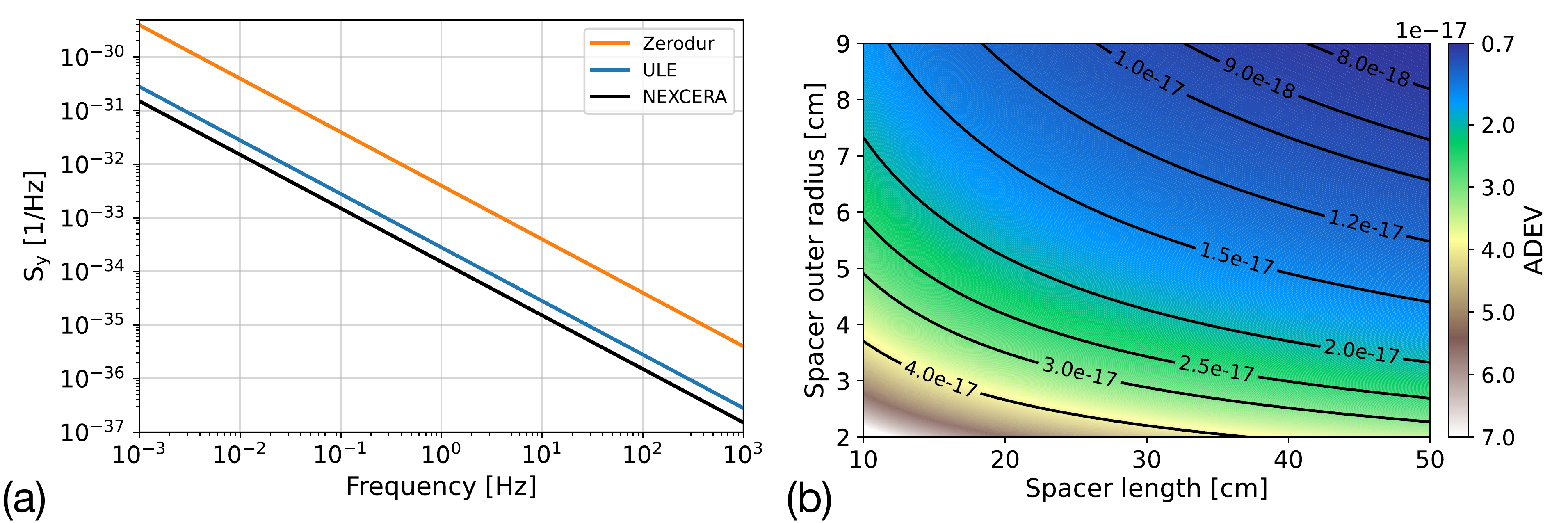}
	\caption{(a) Calculation of the power spectral density of fractional frequency fluctuations $S_y (f)$ of the \SI{30}{\centi\meter} long spacer made of Zerodur, ULE and measured in this work NEXCERA N117B. The outer and inner bore radii of each spacer are set to \SI{5}{\centi\meter} and \SI{6.5}{\milli\meter}, respectively.
	(b) The projected ADEV limit of the NEXCERA spacer in a function of spacer length and spacer outer radius. The inner bore radius is fixed to \SI{6.5}{\milli\meter}, as in Refs.~\cite{Matei_2017, hafner20158}.}
	\label{fig:spacer_noise_both}
\end{figure}

Figure~\ref{fig:spacer_noise_both}(a) illustrates the power spectral density of fractional frequency fluctuations $S_{y}(f)$ of the Brownian thermal noise for a \SI{30}{\centi\meter} long spacer made of ULE, Zerodur, and NEXCERA in the low-frequency regime far from resonance \cite{Kessler_2012} (see Eq.~(S1) and Eq.~(S9) in Supplement~1). The fundamental resonance frequency $f_0$ of each spacer is determined analytically as $f_0 = (2L)^{-1} \sqrt{E/\rho}$, where $L$, $E$, and $\rho$ denote the spacer length, Young's modulus, and density, respectively. For a \SI{30}{\centi\meter} spacer, the first longitudinal resonance mode appears at \SI{9.957}{\kHz} for Zerodur, \SI{9.218}{\kHz} for ULE, and \SI{12.277}{\kHz} for NEXCERA.

The results in Fig.~\ref{fig:spacer_noise_both}(a) show that the Brownian thermal noise of NEXCERA is about a factor of two lower than that of ULE, and also lower than that of Zerodur. This reduction can be attributed to NEXCERA’s higher Young’s modulus (\SI{140}{\GPa}), which is roughly twice that of ULE, while its mechanical loss remains comparable. In comparison with Zerodur, NEXCERA offers both a significantly lower mechanical loss and a higher Young’s modulus, leading to substantially reduced thermal noise. In addition, the resonance frequency of NEXCERA is higher than that of ULE and Zerodur, corresponding to a larger cavity bandwidth.

It is well known that the analytical expression in Eq.~(S1) provides only a rough estimate, as it assumes a uniform stress distribution from the laser beam on the mirror surface~\cite{Kessler_2012}. Despite this simplification, it remains valuable for qualitative comparisons and for assessing the relative thermal noise performance of different spacer materials. Beyond this, realistic evaluations must also account for the way ultra-stable laser cavities are mounted, since carefully optimized mounting configurations are essential to suppress vibration sensitivity and directly affect the thermal noise budget~\cite{Matei_2017,hafner20158}. For these reasons, we retain the analytical expression in the manuscript as a convenient tool for intuitive estimates, while illustrating its limitations through FEM simulations in Supplement~1.

For all the spacer materials presented in Fig.~\ref{fig:spacer_noise_both}(a), the CTE is in the order of $\sim\SI{e-8}{\per\K}$, see Tab.~S1. When operated near the respective zero-crossing temperature, where the effective CTE is minimized, the resulting thermoelastic noise becomes negligible compared to Brownian thermal noise~\cite{Liu_2000, Chalermsongsak_2015}.
However, the slope of the CTE for NEXCERA is significantly steeper than that of ULE~\cite{Ito2017, Hosaka_2013}. This means that tighter temperature stabilization is required, since cavity length changes occur more rapidly in response to temperature fluctuations. In addition, temperature gradients within the spacer—especially in longer cavities—must also be carefully controlled, as they may further limit the achievable performance.

The long-term stability of ultra-stable optical cavities is typically represented in the time domain by the Allan deviation (ADEV)~\cite{Ito2017, hafner20158}.
The ADEV of the Brownian thermal noise is given by a constant value and can be calculated from the power spectral density as follows: ${\sigma_{y} = \sqrt{2\ln(2)S_{y}f}}$~\cite{Kessler_2012}.
In Fig.~\ref{fig:spacer_noise_both}(b), we illustrate the ADEV for the NEXCERA spacer in the shape of a hollow cylinder as a function of spacer length and outer radius.
The inner bore radius is fixed at \SI{6.5}{\milli\meter} as in Refs.~\cite{Matei_2017,hafner20158}.
As the outer radius (and consequently the mass) increases, the thermal noise floor decreases.
This is because the mechanical impedance of the system increases with mass, and thermal noise, according to the fluctuation-dissipation theorem, is inversely proportional to the impedance~\cite{Saulson_1990}.

The calculated ADEV values are several orders of magnitude lower across all spacer lengths and outer radii than the best previously reported NEXCERA cavity, which reached a minimum ADEV of $\sigma_y \approx 10^{-15}$~\cite{Ito2017}. 
Previous studies with NEXCERA spacers and ULE mirrors~\cite{Ito2017, Kwong_2018} focused primarily on temperature sensitivity and long-term drift, rather than on achieving ultra-low fractional frequency instability. 
As a result, the stability reported in Ref.~\cite{Ito2017} was likely limited by technical noise sources rather than by the intrinsic thermal noise of the spacer or mirror coatings.
It is well established that the spacer’s thermal noise contribution is much smaller than that of the mirror substrates and coatings~\cite{Kessler_2012}, and therefore does not limit the overall cavity performance under typical conditions. 
Based on the measured mechanical loss presented here, our estimates confirm that Brownian thermal noise from the spacer can be excluded as the limiting factor for both the previously reported NEXCERA cavities and for the geometries considered in Fig.~\ref{fig:spacer_noise_both}(b).
Only for very short cavities ($<\SI{10}{\milli\meter}$), as may required in some transportable systems, does the spacer’s thermal noise become a significant contribution.

\begin{figure}[tb]
	\centering
	\includegraphics[width=\linewidth]{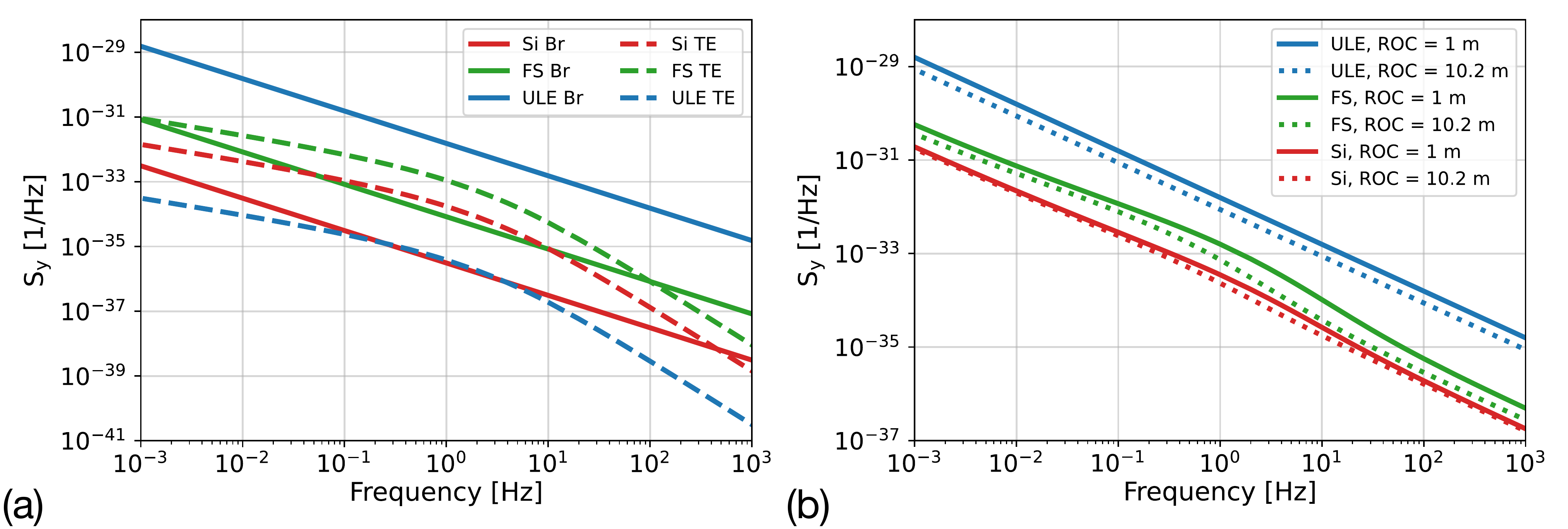}
	\caption{(a) Brownian (Br) and thermoelastic (TE) noise contributions shown as the fractional power spectral density $S_{y}(f)$ for three substrate materials: ULE, fused silica (FS), and silicon (Si). The spacers are \SI{30}{\centi\meter} long, and the mirrors are plano-concave with radii of curvature (ROC) of \SI{1}{\meter}.
    (b) Total $S_{y}(f)$ of a hemispherical NEXCERA spacer cavity for three substrate materials with crystalline coatings, shown for two plano-concave mirror configurations with ROC of \SI{1}{\meter} and \SI{10.2}{\meter}.}
	\label{fig:thermal_noise_both}
\end{figure}

\subsection{Noise performance of NEXCERA spacers with fused silica and silicon mirror substrates}

To evaluate the performance of the entire cavity, including spacer and mirrors, we now analyze the complete system. The best room-temperature optical cavities use ULE spacers with fused silica (FS) mirror substrates~\cite{Matei_2017,hafner20158}. At room temperature, FS exhibits mechanical losses up to two orders of magnitude lower than ULE~\cite{numata2004thermal,Schroeter_2007}, resulting in superior overall cavity performance compared to cavities with ULE substrates.

To date, all reported ultra-stable optical cavities using NEXCERA spacers have employed ULE mirror substrates~\cite{Ito2017, Hosaka_2013, Kwong_2018}. Since NEXCERA and ULE have similar CTEs and zero-crossing temperatures near room temperature~\cite{Hosaka_2013}, combining them does not significantly shift the cavity’s effective zero-crossing temperature.

Figure~\ref{fig:thermal_noise_both}(a) shows the Brownian and thermoelastic noise (see Eq.~(S2) and Eq.~(S3) in Supplement~1) for ULE, FS, and silicon mirror substrates. Like NEXCERA, ULE exhibits negligible thermoelastic noise, but its high mechanical loss makes Brownian noise dominant. Silicon offers the lowest Brownian noise and the smallest total noise at room temperature; however, its CTE~\cite{Middelmann_2015} is about five times larger than that of FS~\cite{Fox_2009}, leading to higher thermoelastic noise. FS, meanwhile, benefits from compensation rings that mitigate CTE mismatch~\cite{Legero_2010}. Additionally, silicon’s opacity below \SI{1}{\micro\meter} limits its use in visible-light applications. Therefore, FS substrates are considered a more versatile choice here.

Figure~\ref{fig:thermal_noise_both}(b) shows the total thermal noise of a hemispherical cavity with a NEXCERA spacer, crystalline coatings~\cite{cole2013tenfold}, and three substrate configurations. Since Brownian and thermoelastic noise depend on the beam spot size, two radii of curvature (ROC) are considered: the commonly used \SI{1}{\meter}~\cite{hafner20158} and the largest commercially available high-finesse ROC of \SI{10.2}{\meter}.

Replacing ULE substrates with FS or silicon improves cavity performance by more than an order of magnitude across the frequency range, measured in power spectral density. This improvement arises from silicon’s low mechanical loss and high stiffness, which benefit both coatings and substrates (see Eq.~(S2), Eq.~(S6), Eq.~(S9), and Tab.~S1 in Supplement~1).

\section{Conclusion}

This work presents the first mechanical loss measurements for NEXCERA N117B, reporting a lowest loss of $\phi = \num{1.89e-5}$, significantly lower than Zerodur ($\phi = \num{3.2e-4}$) and even slightly higher than ULE ($\phi = \num{1.6e-5}$). Based on these results, we calculate the theoretical Brownian thermal noise performance of cavities with NEXCERA spacers, showing it outperforms both ULE and Zerodur. Although the ultimate thermal noise limit of cavities is typically dominated by mirror substrates or coatings, demonstrating that the spacer material itself performs better than ULE represents an important step forward. Together with its lower drift rates and higher stiffness, these results establish NEXCERA as a particularly promising candidate for next-generation interferometric experiments probing fundamental physics, such as tests of general relativity, the equivalence principle, and variations of fundamental constants.

\begin{backmatter}

\bmsection{Funding}
EMPIR programme; European Union's Horizon 2020 research and innovation programme; Deutsche Forschungsgemeinschaft (EXC-2123 QuantumFrontiers—390837967).

\bmsection{Acknowledgment}
The authors sincerely thank S. Häfner, T. Legero, and U. Sterr for their valuable discussions and insights. We also appreciate J. Dickmann for his technical support.
This work was supported by the project 20FUN08 NEXTLASERS, which has received funding from the EMPIR programme co-financed by the Participating States and the European Union's Horizon 2020 research and innovation programme.
Additional funding was provided by the Deutsche Forschungsgemeinschaft (DFG, German Research Foundation) under Germany's Excellence Strategy—EXC-2123 QuantumFrontiers—390837967.

\bmsection{Disclosures}
The authors declare no conflicts of interest.

\bmsection{Data availability}
Data underlying the results presented in this paper are not publicly available at this time but may be obtained from the authors upon reasonable request.

\bmsection{Supplemental document}
See Supplement~1 for supporting content.

\end{backmatter}


\bibliography{bibliography}

\end{document}